# Some New Results on $l_1$-Minimizing Nullspace Kalman Filtering for Remote Sensing Applications

Otmar Loffeld, Center for Sensorsystems, University of Siegen, loffeld@zess.uni-siegen.de, Germany
Dunja Alexandra Hage, Center for Sensorsystems, University of Siegen, hage@zess.uni-siegen.de, Germany
Miguel Heredia Conde, Center for Sensorsystems, University of Siegen, heredia@zess.uni-siegen.de, Germany
Ling Wang, Key Lab. of Radar Imaging and Microwave Photonics, Nanjing University of Aeronautics and Astronautics, lingnuaa@zess.uni-siegen.de, China

## Abstract

The SAR image generation from raw data can be seen as a decoding problem of raw data which has been encoded by the SAR sensor, where the coding scheme depends on the sensor and the sensing geometry. Compressed sensing (CS) decoding approaches, implicitly model the forward encoding or sensing process in matrix vector form with a sensing matrix, where the dimensions of this matrix grow proportionally with the product of raw data samples and finally wanted image pixels. Especially when applied to Synthetic Aperture Imaging Processing we then face the necessity of handling extremely large data vectors and even larger sensing matrices. Classical focussing/decoding approaches conceptually invert the system of equations, introducing matched filter concepts making use of pseudoinverses. CS concepts under the assumption of images which are sparse in some basis or in combinations of bases promise a reduction of raw data samples without sacrificing resolution or vice versa an improved resolution or signal to noise ratio if the number of raw data samples is not reduced. Consequently the Cramer Rao Bound assuming sparsity of the image to be reconstructed is governed by the number of non-zero coefficients in its sparse description which under an oracle given knowledge of the sparse support converts the forward sensor model into an even overdetermined system of equations which would allow a smaller Cramer Rao Bound than without sparsity assumption.

Basis pursuit approaches to CS decoding employ a constrained $l_1$-minimization for the wanted (sparse) image vector to be reconstructed where the constraining data fidelity term models the SAR sensor's encoding, which requires a burdensome matrix vector multiplication in each iteration step. Hence algorithmic issues in realizing the CS decoding play a crucial and decisive role. This paper presents some new results on recursive $l_1$-minimization by Kalman filtering. We consider the $l_1$-norm as an explicit constraint, formulated as a nonlinear $l_1$-observation of the state to be estimated. Interpreting a sparse vector to be estimated as a state which is observed from erroneous (undersampled) measurements we can address time- and space-variant sparsity, any kind of *a priori* information and also easily address nonstationary error influences in the measurements available. Rather than estimating the full $n$-dimensional vector of unknowns, the presented approach works on a nullspace basis and only estimates the nullspace vector of dimension $n$-$m$, where $m$ is the dimension of the observation vector. Due to this and depending on $m$, the computational burden of this nullspace vector estimation approach may be considerably smaller in comparison with Chambolle and Pock's Primal Dual Algorithm or many other algorithm estimating the full $n$-dimensional vector of unknowns. In fact for mildly undetermined systems of equations the computational complexity of the Nullspace Kalman Filter can beat Chambolle and Pock's primal dual algorithm and approaches the OMP complexity, which depends on the sparsity, whereas it does not for the Nullspace Kalman filter.

## 1 Introduction

Conceptually, the reconstruction of a sparse vector from an underdetermined system of equations can be achieved as the solution of an $l_1$-minimization problem subject to a data fidelity constraint by linear programming, which provides the correct solution and is computationally feasible. Recent approaches like, e.g., Chambolle and Pock's primal dual optimization algorithm [1] (C&P), have not only demonstrated to efficiently solve this problem, but also to exhibit optimal convergence $O(1/n_i^2)$ over the number of iterations $n_i$ for convex problems and hence can be interpreted as a benchmarking algorithm for $l_1$-minimization problems with convex data fidelity terms.

On the other hand and from an estimation theoretical point of view, the optimal reconstruction of a vector of dynamically changing unknowns from linear observations in a noisy Gaussian environment leads to Kalman filtering concepts, which conceptually can handle the minimal $l_1$-norm as an additional constraint in form of one additional measurement, which is iteratively minimized, yielding an optimal estimate under sparsity con-ditions. While this approach is not completely new, the approach described in [2, 3], rather than estimating the full n-dimensional vector of unknowns only estimates the nullspace vector of dimension $n-m$, where $m$ is the dimension of the observation vector. Due to this and depending on $m$ the computational burden of this nullspace vector estimation approach may be considerably smaller in comparison with C&P's or any other algorithm estimating the full n-dimensional vector of unknowns. In fact for mildly undetermined systems of equations the computational efficiency of the Nullspace Kalman Filter beats C&P's algorithm approaching the OMP complexity, which depends on the sparsity, whereas it does not for the Nullspace Kalman filter. The paper will also demonstrate by Donoho-Tanner graphs that the Nullspace Kalman Filter and C&P's algorithm asymptotically perform identically concerning reconstruction error, but with completely complementary efficiency characteristics: While the computational burden increases with increasing number of measurements for C&P's primal dual approach, it decreases for the Nullspace Kalman Filter.

The rest of the paper is organized as follows. In sections 2 and 3, we will introduce some notation, then shortly review the role of the nullspace and revisit the Nullspace Kalman filter approach in condensed form. We will also introduce some ideas in section 3 for improving the convergence rate, based on Aitken's convergence acceleration, which are essential for the Kalman filter to achieve competitive results. The first subsection of Section 4 will then compare the Nullspace Kalman Filter with the C&P approach using an (optimal in terms of coherence) sensing matrix, while the second subsection will present a small but descriptive set of results for Inverse SAR Imaging (ISAR), where the Nullspace Kalman filter is compared with the Range Doppler approach, the Primal/Dual Approach, a greedy OMP solution and a greedy Kalman filter implementation of an OMP. In section 5 we draw a few conclusions.

## 2 Notation

The $l_p$-norm in $\mathbb{C}^n$ is denoted by $\|\cdot\|_p$. The support of $\underline{x}$ is denoted as $\mathrm{supp}(\underline{x}) := \{i : x_i \neq 0\}$ and $\|\underline{x}\|_0 = |\mathrm{supp}(\underline{x})|$ is the number of non-zero elements of $\underline{x}$, where $|\cdot|$ denotes the cardinality of a set. A set of difference vectors is given as: $\mathcal{F} - \hat{\underline{x}} := \{\underline{x} - \hat{\underline{x}} : \underline{x} \in \mathcal{F}\}$, $\mathcal{F} \subset \mathbb{C}^n$ and $\hat{\underline{x}} \in \mathbb{C}^n$.

## 3 Nullspace and Nullspace-based $l_1$-minimizing Kalman Filter

### 3.1 LQ Factorization and Weighted Least Squares Estimation

We adopt the notation of classical estimation theory, where a vectorial observation $\underline{y} \in \mathbb{C}^m$ of some state $\underline{x} \in \mathbb{C}^n$ is given by an observation matrix: $C \in \mathbb{C}^{m \times n}$ as
$$\underline{y} = C\underline{x} + \underline{\eta} \qquad 3.1$$
η is assumed as zero-mean Gaussian noise with covariance R. If $m<n$ we have an underdetermined system of equations for which a manifold of solutions exists. The nullspace of the matrix $C$ is denoted by:
$$\mathcal{N}(C) := \{\underline{x} : C\underline{x} = \underline{0}\} \qquad 3.2$$
and can be calculated from the LQ factorization:
$$C = L \cdot Q \qquad 3.3$$
$L = \begin{bmatrix} L_1 & 0_{m \times (n-m)} \end{bmatrix}$ is a left triangular matrix, while $Q \in \mathbb{C}^{n \times n}$ is an unitary matrix $Q^H \cdot Q = I$, decomposed as:
$$Q = \begin{bmatrix} Q_1 \\ Q_2 \end{bmatrix} \quad Q_1 \in \mathbb{C}^{m \times n}, Q_2 \in \mathbb{C}^{(n-m) \times n} \qquad 3.4$$
where the $n-m$ row vectors of $Q_2$ are the $n$-dimensional nullspace vectors of $C$. In [2, 3] it was shown that a particular (non-unique) solution of 3.1, which is the LS estimate, can be obtained as:
$$\hat{\underline{x}}_P = Q_1^H L_1^{-1} \underline{y} \qquad 3.5$$
where any nullspace content of $\underline{x}$ has got lost. Now denoting the manifold of solutions to the weighted least squares problem by:
$$\mathcal{F}\{\underline{y}\} := \left\{\underline{x} : \mathrm{E}\left[\left[\underline{y} - C\underline{x}\right]^H R^{-1} \left[\underline{y} - C\underline{x}\right]\right] = \min\right\} \qquad 3.6$$
it has been shown in[3] that the manifold of solutions can be decomposed as:
$$\mathcal{F}\{\underline{y}\} = \mathcal{N}(C) + \hat{\underline{x}}_P \qquad 3.7$$
Equation 3.7 indicates that the set of all difference vectors between the manifold of feasible solutions and a specific estimate (particular solution) is the nullspace:
$$\mathcal{F}\{\underline{y}\} - \hat{\underline{x}}_P = \mathcal{N}(C) \qquad 3.8$$
Using the assumption of sparsity on $\underline{x} \in \mathbb{C}^n$ $\|\underline{x}\|_0 = |\mathrm{supp}(\underline{x})| = s \leq m$, we formulate the reconstruction problem as follows:
$$\hat{\underline{x}}_{opt} = \arg\min_{\hat{\underline{x}} \in \mathcal{F}\{\underline{y}\}} \|\hat{\underline{x}}\|_1 \qquad 3.9$$
Substituting $\mathcal{F}\{\underline{y}\}$ by the last equality of 3.7 we obtain:
$$\hat{\underline{x}}_{opt} = \hat{\underline{x}}_P + Q_2^H \arg\min_{\underline{x}_\nu} \|\hat{\underline{x}}_P + Q_2^H \underline{x}_\nu\|_1 \qquad 3.10$$
$\underline{x}_\nu$ is the minimal $(n-m)$ representation of the nullspace vector $\underline{v}_x$ while the weighted least squares estimate:
$$\hat{\underline{x}}_P = Q_1^H L_1^{-1} \underline{y} \qquad 3.11$$
can be calculated offline or in a first step from the available measurements in a first step. Due to the underdetermined system of equations this estimate is not unique. Then we seek for that specific nullspace vector which adds up to the least squares solution in a way that the overall $l_1$-norm of the sum is minimized: This reflects the sparse nature of the vector we are looking for. Furthermore equation 3.10 instructs us that rather than seeking the $n$-dimensional nullspace vector $\underline{v}_x(k)$, we only need to estimate its ($n$-$m$)-dimensional (minimal) representation $\underline{x}_\nu$.

## 3.2 Nullspace based extended linearized Kalman Filter

The Kalman Filter given in [2] only estimates the nullspace vector in its minimal formulation. In predictor-corrector structure, the Kalman filter calculates a prediction estimate and its corresponding error covariance for step $k+1$ based on an available filter estimate at $k$ by:

$$\hat{\underline{x}}_\nu^-(k+1) = \hat{\underline{x}}_\nu^+(k); \quad P_\nu^-(k+1) = P_\nu^+(k) + Q(k) \quad 3.12$$

Due to the constant nature of the state, the prediction for iteration at step $k+1$ is the same as the filtered state estimate at $k$, only the uncertainty increases. Then the EKF evaluates the Jacobian matrix of the nonlinear observation in the prediction point by:

$$C_\nu(k+1) = \frac{dh(\underline{x})}{d\underline{x}}\bigg|_{\hat{\underline{x}}_P + E_N \hat{\underline{x}}_\nu^-(k+1)} \cdot E_N = H(k+1) \cdot E_N \quad 3.13$$

where we have introduced:

$$E_N = Q_2^H$$

$$H(k+1) = \begin{bmatrix} \frac{dh(\underline{x})}{dx_1} & \frac{dh(\underline{x})}{dx_2} & \cdots & \frac{dh(\underline{x})}{dx_n} \end{bmatrix}\bigg|_{\hat{\underline{x}}_P + E_N \hat{\underline{x}}_\nu^+(k)} \quad 3.14$$

The Kalman gain matrix is evaluated:

$$K(k+1) = P_\nu^-(k+1)C_\nu^H(k+1) \cdot \underbrace{\left[C_\nu(k+1)P_\nu^-(k+1)C_\nu^H(k+1) + r(k+1)\right]^{-1}}_{scalar} \quad 3.15$$

$r(k+1)$ is the scalar noise covariance of the $l_1$-norm observation. In each iteration we force the $l_1$-norm to decrease by letting the new observation be smaller than the last $l_1$-norm of the overall estimate:

$$y(k+1) = \gamma_{k+1} l_{emp} = \gamma_{k+1} \left\| \hat{\underline{x}}_P + E_N \hat{\underline{x}}_\nu^+(k) \right\|_1$$
$$= \gamma_{k+1} \left\| \hat{\underline{x}}^+(k+1) \right\|_1 \quad 3.16$$

Then the prediction estimate is corrected to the filtered state estimate by adding the residual between new (decreased) and the preceding $l_1$-norm

$$\hat{\underline{x}}_\nu^+(k+1) = \hat{\underline{x}}_\nu^-(k+1) + K(k+1)\left[y(k+1) - h(\hat{\underline{x}}_\nu^-(k+1))\right] \quad 3.17$$

Finally the error covariance of that estimate is calculated and the empirical $l_1$-norm of the overall estimate is determined by:

$$P_\nu^+(k+1) = P_\nu^-(k+1) - K(k+1)C_\nu(k+1)P_\nu^-(k+1)$$
$$l_{emp} = \left\| \hat{\underline{x}}_P + E_N \hat{\underline{x}}_\nu^+(k+1) \right\|_1 = \left\| \hat{\underline{x}}^+(k+1) \right\|_1 \quad 3.18$$

and the new iteration cycle can begin with equation 3.12. In order to start the Kalman filter we need to have calculated $\hat{\underline{x}}_P$ by 3.5 and the initial values

$$\hat{\underline{x}}_\nu^+(0) = \underline{0}; \quad P_\nu^+(0) = 0_{(n-m)\times(n-m)}; \quad l_{emp} = \gamma_1 \left\| \hat{\underline{x}}_P \right\|_1 \quad 3.19$$

## 3.3 Convergence Acceleration

Aitkens $\Delta^2$-basic process [4] is a convergence-accelerated method for iterative staggered solution processes. If the sequence of consideration is similar enough to a geometric sequence, the process leads to an acceleration of the convergence. First the accelerated algorithm is started with the initialization by 3.19. The parameter $\gamma(k+1)$ in 3.16 is redefined as $\gamma(k+1) = 1 - \tilde{r}(k+1)$ with $\tilde{r} \in \mathbb{R}$. In the second iteration step the measurement $y$ is determined by a relaxation parameter $\omega$ after the modified extrapolation method according to Aitken's delta-squared process

$$y(k) = -\tilde{r}(k)\left(\|\hat{\underline{x}}(k)\|_1 + \omega\left(\|\hat{\underline{x}}(k)\|_1 - \|\hat{\underline{x}}(k-1)\|_1\right)\right)$$
$$3.20$$

In the third iteration step the measurement $y$ is calculated with using Steffensen's method [5]

$$y(k) = \frac{y(k)y(k-2) - (y(k-1))^2}{y(k) - 2y(k-1) + y(k-2)} \quad 3.21$$

From the third iteration step the coefficient $\tilde{r}$ is obtained via Steffensen's method in a converging sequence $\tilde{r}(k) = (1 - \hat{r}(k))\tilde{r}^-(k)$, where $\hat{r}$ is actually calculated using Steffensen's method. Furthermore Steffensen's method makes $\tilde{r}(k)$ tend to zero when $k \to \infty$. The value $y$ converges to a limit value different from zero.

## 4 Performance Analysis and Results

### 4.1 Comparison with Primal Dual Algorithm

For each experiment, an $s$-sparse signal $\underline{x} \in \mathbb{C}^n$ is generated at random. Both the real and imaginary parts of each nonzero complex coefficient were drawn from *i.i.d.* normal distributions of zero mean and unit variance, and the resulting $\underline{x}$ is then $l_2$- normalized.

We used *best complex antipodal spherical codes* (BCASC) as measurement matrix $C \in \mathbb{C}^{m \times n}$ for obtaining the vector of measurements $\underline{y} \in \mathbb{C}^m$. We consider different experimental cases for different values of the parameters $\delta = m/n$ and $\rho = s/m$. More specifically, we evaluate the entire $\delta - \rho$ plane, i.e., $0 \leq \delta \leq 1, 0 \leq \rho \leq 1$ by means of 16 equally-spaced discrete steps per parameter. For all experiments the signal length is set to $n = 128$. The performance of the different alternatives is evaluated in terms of the $l_2$ recovery error $\|\underline{x} - \hat{\underline{x}}\|_2$.

The comparison of the empirical behaviour of Chambolle and Pock's primal dual algorithm with the Nullspace Kalman Filter reveals a striking and surprising similarity, both algorithms despite of their completely different backgrounds show almost identical results concerning $l_2$ recovery error. With increasing observation vector dimension the Kalman Filter performs faster, while all other approaches become slower.

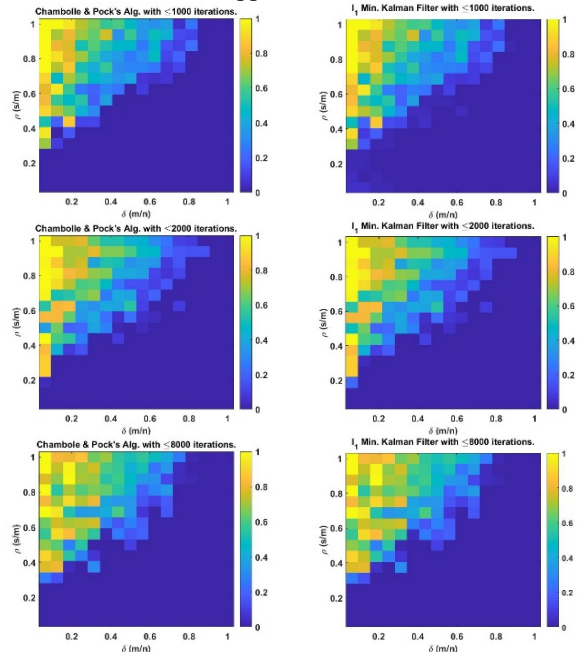

Figure 1: Donoho-Tanner Graphs comparing the reconstruction error of Chambolle and Pocks's primal/dual algorithm (left) with Nullspace Kalman Filter (right).

## 4.2 Empirical results for ISAR Imaging

We used an ISAR aircraft data set to demonstrate the performance of the nullspace $l_1$ minimization image reconstruction. The dimension of the image in range and azimuth was $N_r \times N_a = 100 \times 80$ where the number of measurements $m$ was reduced to 4047 (approx. 50%) and 2000 (25%). The sparsity s estimated by the approach in [3] was 900. Note that all ISAR data sets (data 1 and data 2) were motion compensated by a global entropy minimization based range alignment method [6]. The measurement reduction was obtained by random downsampling in both range and cross range dimension. As for the parameters of the KF, we set $P_{init} = 0$ and the covariance matrices $Q_k$ and $R_k$ both to be an identity matrix. In our settings, the prior estimates and the measurements contribute to the reconstruction to the same extent, which ensures better reconstruction performance as compared to the cases in which the estimation either only relies on the measurements or only on the preceding estimates.

### 4.2.1 Aircraft image reconstruction with 25% measurements.

Fig. 2 presents the reconstructed images of aircraft data using 2000 measurements. For performance comparison, we present the imaging results using the conventional RD approach, as shown in Fig. 2(a), OMP, as shown in Fig. 2(b) and Chambolle-Pock's Primal-Dual $l_1$-minimization [1], as shown in Fig. 2(c). Fig. 2(d) shows the image reconstructed by the Nullspace $l_1$-minimization KF approach. All images presented in this section are displayed with the same contour levels. Comparing Fig. 2(d) with Fig. 2(b), we see the Nullspace $l_1$ minimization KF provides better reconstruction than OMP. There are fewer artifacts in the images reconstructed by the nullspace $l_1$-minimization KF. There are more artificial points reconstructed in Fig. 2(c) than there are in 2(d).

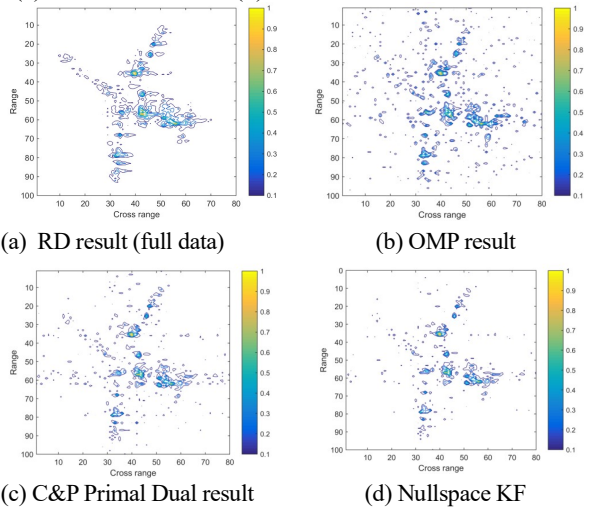

(a) RD result (full data)  (b) OMP result

(c) C&P Primal Dual result  (d) Nullspace KF

Figure 2: Imaging results of aircraft data 1 using 25% measurements. (a) RD image obtained using full data; (b) OMP result; (c) C&P Primal-Dual result; (d) Nullspace $l_1$-minimization KF

### 4.2.2 Numerical Evaluation

We consider the RD image as a reference for evaluating the numbers of "False Alarms" (*FA*), "Missed Detections" (*MD*) and the "Relative Root Mean Square Error" (*RRMSE*) given as:

$$RRMSE = \sqrt{\frac{1}{K}\sum_{T_k=1}^{K}\left(\frac{|A_T(T_k)| - |A_{CS}(T_k)|}{|A_T(T_k)|}\right)^2} \quad 4.1$$

where $K$ is the total number of scatterers, $T_k$ denotes the index of the scatterer in the image, $A_T$ and $A_{CS}$ are the vectors containing the complex amplitude of the scatterer in the RD image and CS reconstructed image. The larger the value of the *RRMSE*, the greater the brightness difference between the reconstructed image and the RD image. We introduce the "Target-to-Clutter Ratio" (*TCR*) :

$$TCR = 10\log_{10}\left(\frac{\frac{1}{N_T}\sum_{(i,j)\in T}|\hat{\mathbf{I}}(i,j)|^2}{\frac{1}{N_{Clu}}\sum_{(i,j)\in Clu}|\hat{\mathbf{I}}(i,j)|^2}\right) \quad 4.2$$

where $(i, j)$ represents the pixel index in the image, $\hat{\mathbf{I}}$ represents the image, $\hat{\mathbf{I}}(i,j)$ denotes the reconstructed value of the pixel at the position $(i, j)$ of the recon-structed image, $T$ represents the target region, $Clu$ represents the clutter region in the image, $N_T$ represents the number of pixels in the target region and $N_{Clu}$ represents the number of pixels in the clutter region. The larger *TCR* is, the better the reconstructed image quality is. The "Image Entropy" (*IE*) is introduced as:

$$IE = \sum_{i=1}^{Nr}\sum_{j=1}^{Na} \frac{|\hat{\mathbf{I}}(i,j)|^2}{\Sigma_{\hat{\mathbf{I}}(i,j)}} \ln \frac{\Sigma_{\hat{\mathbf{I}}(i,j)}}{|\hat{\mathbf{I}}(i,j)|^2} \qquad 4.3$$

where: $\Sigma_{\hat{I}(i,j)} = \sum_{i=1}^{Nr}\sum_{j=1}^{Na}|\hat{\mathbf{I}}(i,j)|^2 \qquad 4.4$

and the "Image Contrast" (*IC*) is given as:

$$IC = \frac{\sqrt{E[(|\hat{\mathbf{I}}|^2 - E[|\hat{\mathbf{I}}|^2])^2]}}{E[|\hat{\mathbf{I}}|^2]} \qquad 4.5$$

where *E* denotes the spatial mean operator. Both image entropy and image contrast measure the image sharpness or focusing quality. A sharper image means better focusing, leading to larger image contrast and smaller image entropy. However, they do account for the artifacts or pseudo scatterers as well as missing scatterers associated with the CS reconstruction. The image reconstruction quality is assessed using these parameters. The results shown in red and green colours in Table 4.1 indicate the best and second best performances.

| Algorithm | FA | MD | RRMSE | TCR(dB) | IE | IC |
|---|---|---|---|---|---|---|
| RD | | | | 18,14 | 5,39 | 8,17 |
| GKF[7] | 145 | 197 | 0,30 | 20,19 | 5,32 | 8,25 |
| Nullspace KF | 76 | 242 | 0.28 | 22,63 | 4,85 | 10,66 |
| OMP | 455 | 221 | 0,65 | 15,19 | 5,93 | 6,07 |
| Primal-Dual | 163 | 210 | 0,28 | 19,85 | 5,21 | 9,17 |

Table 4.1: Image quality assessment for the imaging of an aircraft target using 25% raw data

#### 4.2.3 Aircraft image reconstruction with 50% measurements.

In Fig.3, we see that as the number of the measurements increases, all algorithms provide better reconstructions than in the case of fewer measurements. The nullspace l1 minimizing KF algorithm leads to the smallest artifacts as compared to the Primal-Dual and OMP algorithms.

Figure 3c additionally shows the results of a greedy Kalman filter OMP (GKF)[7] implementation which performs slightly better than the OMP algorithm. It is worth noting that with increasing number of measurements, the computational time of Primal-Dual, OMP and GKF increase as expected, but the Nullspace KF becomes much faster and the computational time is of the same order of magnitude as the OMP algorithm. This is due to the reduction of the nullspace dimension for a larger number of measurements.

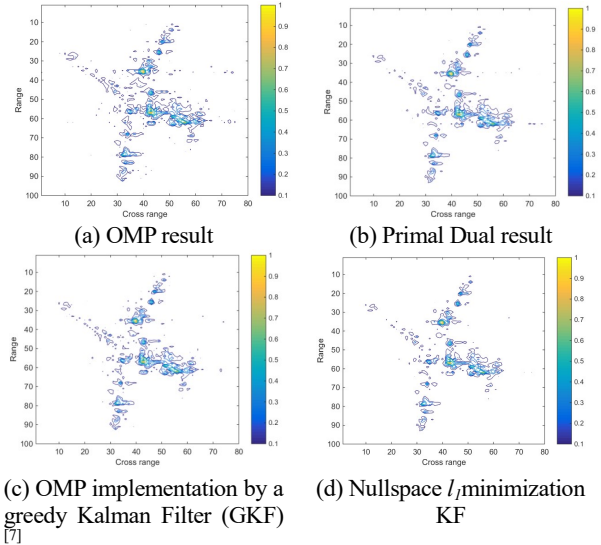

(a) OMP result  (b) Primal Dual result
(c) OMP implementation by a greedy Kalman Filter (GKF)[7]  (d) Nullspace $l_1$ minimization KF

Figure 3: Imaging results of aircraft data 1 using 50% measurements. (a) OMP; (b) Primal-Dual; (c) GKF; (d) Nullspace $l_1$-minimizing KF.

## 5 Conclusions

The proposed nullspace $l_1$-minimization method empirically yields the best image reconstruction performance and turns out very robust. In most cases, Chambolle-Pock's $l_1$- minimization approach provides the second best imaging results, but it is not as robust as other approaches, and in some cases leads to bad reconstructions, The OMP approach works the fastest, but it is inferior to other ones in the image reconstruction quality.

## Acknowledgements

The works of this paper have been funded by DFG, grant number Lo 455/20-1, which is gratefully appreciated.